\documentclass[aps,prl,reprint,superscriptaddress]{revtex4-1}
\usepackage{graphicx}
\usepackage{color}
\usepackage{multirow}
\usepackage{amsmath}
\usepackage{enumitem}
\usepackage{amssymb}
\usepackage{hyperref}

\begin{document}


\title{A generalized approach for rapid entropy calculation of liquids and solids}


\author{Qi-Jun Hong}
\email[Corresponding author:]{qijun.hong@asu.edu}
\affiliation{School for Engineering of Matter, Transport, and Energy, Arizona State University, Tempe, Arizona 85285, USA}

\author{Zi-Kui Liu}
\affiliation{Department of Materials Science and Engineering, Pennsylvania State University, State
College, PA 16802, USA}


\date{\today}

\begin{abstract}
We build a comprehensive methodology for the fast computation of entropy across both solid and liquid phases. The proposed method utilizes a single trajectory of molecular dynamics (MD) to facilitate the calculation of entropy, which is composed of three components. The electronic entropy is determined through the temporal average acquired from density functional theory (DFT) MD simulations. The vibrational entropy, typically the predominant contributor to the total entropy, even within the liquid state, is evaluated by computing the phonon density of states via the velocity auto-correlation function. The most arduous component to quantify, the configurational entropy, is assessed by probability analysis of the local structural arrangement and atomic distribution. We illustrate, through a variety of examples, that this method is both a versatile and valid technique for characterizing the thermodynamic states of both solids and liquids. Furthermore, this method is employed to expedite the calculation of melting temperatures, demonstrating its practical utility in computational thermodynamics.
\pacs{2}


\end{abstract}



\maketitle



Entropy is a fundamental concept in thermodynamics, representing the degree of disorder or randomness in a system. While crucial, accurately computing entropy is challenging due to its non-intuitive nature and the complexity of accounting for the microstates of a system. This computational difficulty underscores the intricacies of predicting system behavior where energy distribution and disorder are intricate variables.

Recent advancements in the calculation of entropy have significantly enhanced our understanding of microstate analysis and energy distribution. These sophisticated methodologies, while providing comprehensive insights into the fundamental aspects of entropy, introduce specific computational challenges and are confined to a narrow application spectrum. The Particle Insertion Method \cite{widom1963,Widom1982} relies on probabilistic evaluations, deeply intertwined with statistical mechanics principles and chemical potential. Despite its efficacy \cite{Hong2012}, it demands significant computational investment, relying on a large amount of insertion trials to achieve statistically significant results, rendering it extremely expensive when integrated with DFT \cite{Hong2012}. The Pair Distribution Function method \cite{green1952molecular, wallace1987density, Widom2019} analyzes spatial particle correlations against an ideal gas benchmark, translating these into entropy metrics. The method, however, relies on an expansion of entropy in multiparticle correlations beyond the pair-entropy term, potentially increasing complexity and resulting in truncation errors and large fluctuations. Lastly, the Two-Phase Thermodynamics Method \cite{Lin2003} presents a creative approach by conceiving the system as a combination of solid-like and gas-like phases. This method uniquely separates solid-state phonon contributions, which possess a well-defined analytical entropy expression, and approximates the remainder of the system as a gas phase, e.g., modeled using hard spheres. The efficacy and accuracy of this method are contingent upon the artificial division into solid-like and gas-like components and the specific selection of the gas-like phase model.

In the present study, we introduce a comprehensive methodology designed to generalize the computation of entropy in both solid and liquid states, especially for high temperature applications, enabling us to calculate entropy from one DFT MD trajectory. The method adopts recent advancements in the field, including the concepts of entropy ``coarse-graining" \cite{vandeWalle2002} and the Zentropy theory \cite{Liu2022Zentropy}, which further segment entropy into distinct categories - configurational, vibrational, and electronic, aiming to encapsulate entropy at various levels, acknowledging its multi-layered nature. (There is no rotational entropy for bulk solids or liquids, as it is part of the vibrational entropy.)
\begin{equation}\label{Equation:compoents}
S = S_{\text{conf}} + S_{\text{vib}} + S_{\text{elec}}.
\end{equation}
These theoretical frameworks offer a granular understanding of free energy and entropy, enabling feasible computation and a more detailed analysis of thermodynamic properties. Furthermore, incorporating $S_{\text{vib}}$ (based on phonon analysis under the harmonic approximation at high temperature) and $S_{\text{elec}}$ (based on quantum mechanical calculation of electronic density of states) allows the assimilation of quantum mechanical effects, such as quantum harmonic oscillation of phonons and the Fermi-Dirac distribution of electrons, thereby facilitating a quantum thermodynamic description that reconciles with classical mechanics at sufficiently high temperatures.

\begin{figure}
    \centering
    \includegraphics[width=0.49\textwidth]{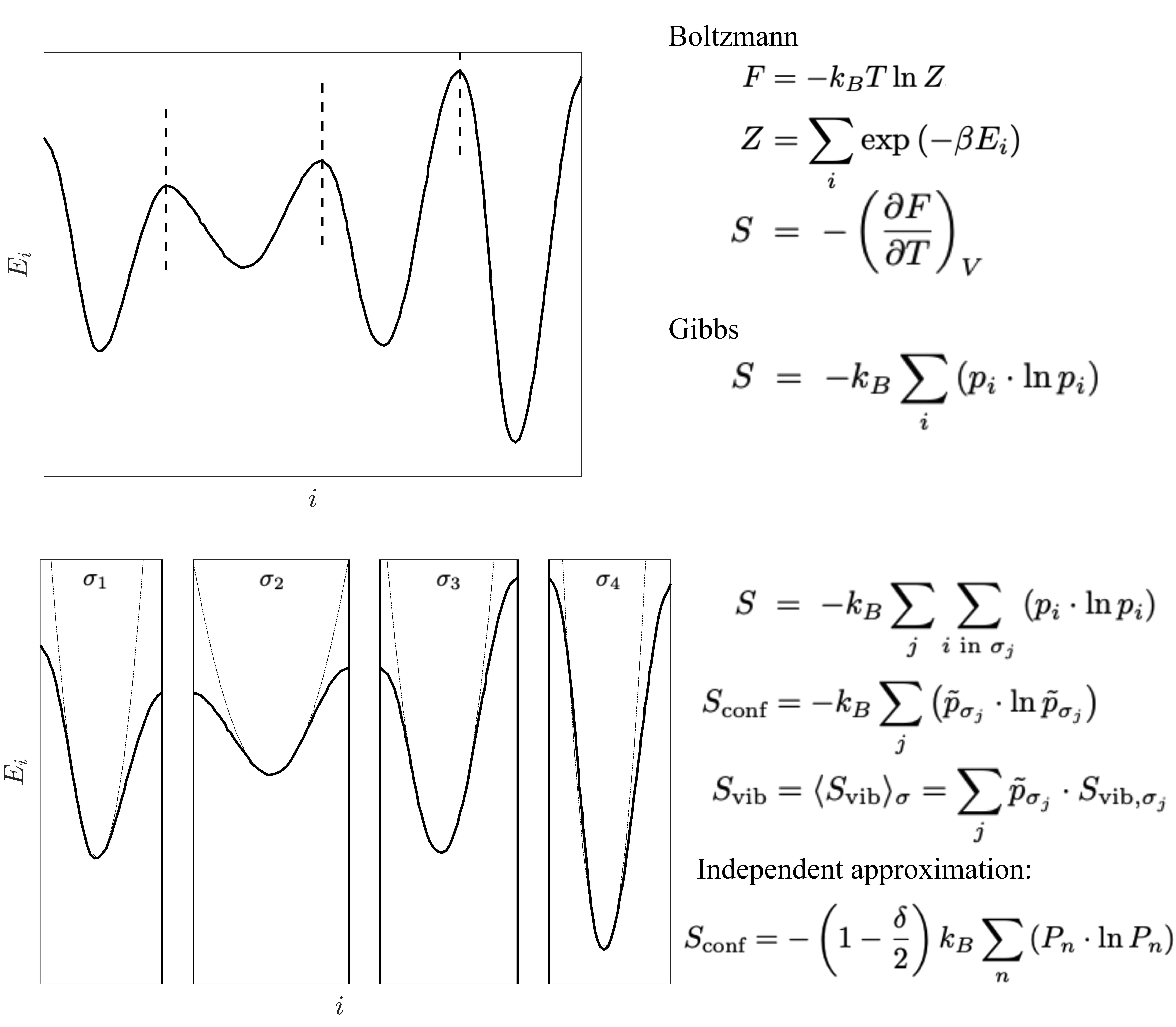} 
    \caption{\label{Fig:diagram}Diagram for segmentation of entropy into configurational and vibrational terms.}
\end{figure}

Here we outline a theory underpinning our methodology, with comprehensive derivations detailed in the Supplemental Material. This theory is anchored in the seminal interpretations of entropy by Boltzmann \cite{perrot1998thermodynamics} and Gibbs et al. \cite{Gibbs1960,vonNeumann1927,shannon1948}, which are demonstrated to be equivalent (see Supplemental Material), thus allowing for their interchangeable application. As shown in Fig. \ref{Fig:diagram}, we write the combined configurational and vibrational entropy ($S_{\text{conf}} + S_{\text{vib}}$) employing Gibbs's formula,
\begin{equation}\label{Eqn:Gibbs_main}
S = -k_B \sum_i{ \left( p_i \cdot \ln{p_i} \right) },
\end{equation}
where $p_i$ denotes the probability of microstate $i$ within the coordinate space $\mathbb{R}$. Proceeding to segment the space into configurations $\sigma$, we separate $S_{\text{conf}}$ and $S_{\text{vib}}$, as detailed in the Supplemental Material,
\begin{eqnarray}
S_{\text{conf}} &=& -k_B \sum_j{ \left( \tilde{p}_{\sigma_j} \cdot \ln{\tilde{p}_{\sigma_j}} \right) }, \label{Eqn:Sconf_main}\\
S_{\text{vib}} &=& \langle S_{\text{vib}} \rangle_{\sigma} = \sum_j{ \tilde{p}_{\sigma_j} \cdot S_{\text{vib},\sigma_j} }\label{Eqn:final_Svib_main},
\end{eqnarray}
where $\tilde{p}_{\sigma_j}$ is the aggregated $p_i$ within configuration $\sigma_j$ and thus the configuration's probability. 
We prove in the Supplemental Material that, under an assumption of atomic probability independence, Eqn. \ref{Eqn:Sconf_main} can be simplified, allowing the substitution of the configuration probability $\tilde{p}_{\sigma}$ with atomic local probability $P_{n}$,
\begin{equation}\label{Eqn:final_Sconf_main}
S_{\text{conf}} = -\left(1-\frac{\delta}{2}\right)k_B \sum_{n}{ \left( P_{n} \cdot \ln{P_{n}} \right) },
\end{equation}
where $P_{n}$ stands for the overall probability of an atom having $n$ nearest neighbors, and $\delta$ is assigned a value of 1 if the pair consists of the same element, and 0 otherwise. For detailed derivations of Eqns. \ref{Eqn:Gibbs_main} through \ref{Eqn:final_Sconf_main}, readers are directed to the Supplemental Material.

\begin{figure}[]
    \centering
    \includegraphics[width=0.45\textwidth]{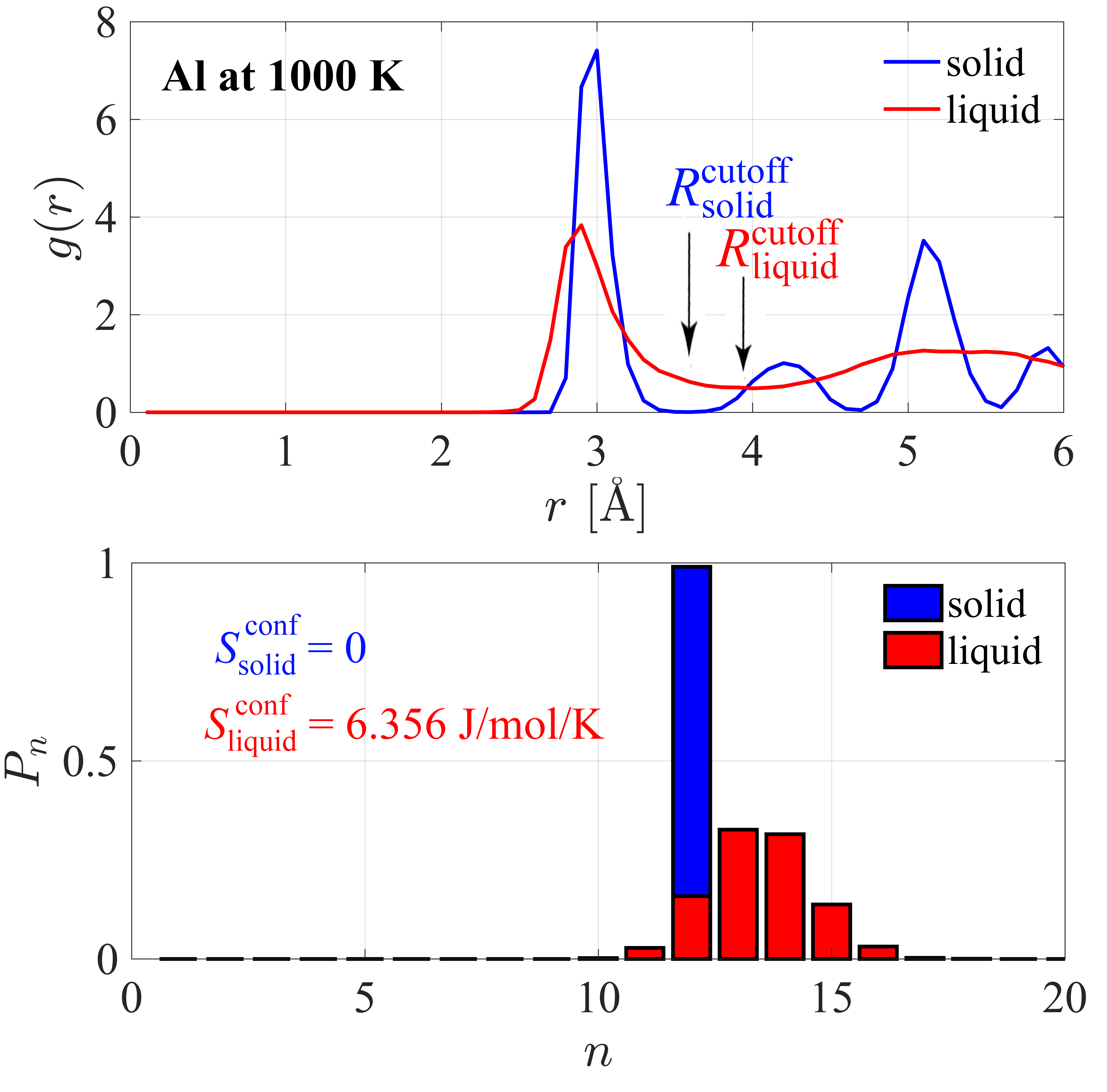} 
    \caption{Radial distribution function (upper) and probability $P_n$ of local configurations of solid (fcc) and liquid aluminum at 1000 K.}
    \label{Figures:conf}
\end{figure}

Utilizing Eqns. \ref{Eqn:final_Sconf_main} and \ref{Eqn:final_Svib_main}, we are now able to compute $S_{\text{conf}}$ and $S_{\text{vib}}$ from a single MD trajectory. The configurational entropy, previously a formidable task, now can be evaluated from the atomic local probability $P_{n}$, the probability of occurrence of a local structural arrangement. To implement this, we perform a pairwise radial distribution function analysis to determine the nearest neighbor cutoff radius. Subsequently, we count the nearest neighbors located within this radial boundary. The resultant probability $P_{n}$ is then applied in Eqn. \ref{Eqn:final_Sconf_main} to compute the configurational entropy, as shown in Fig. \ref{Figures:conf}.

Concurrently, the vibrational entropy $S_{\text{vib}}$ is deduced from the same MD trajectory. According to Eqn. \ref{Eqn:final_Svib_main}, the overall vibrational entropy is the ensemble average of all configurations $\sigma$, attainable through MD's effective configuration sampling, given a sufficiently high diffusion rate. This characteristic is typically observed in liquids or disordered solid states with light elements at high temperatures, where the dynamic nature of the system promotes rapid exploration of configurational space. In this analysis, we pivot to Boltzmann's entropy interpretation within the harmonic approximation (also see the Supplemental Material),
\begin{equation}\label{Eqn:Svib_solid}
F = k_BT\int{ \ln{\left[2\sinh(\frac{h \nu}{2k_BT})\right]}g(\nu)d\nu},
\end{equation}
where $\nu$ is phonon frequency and $g(\nu)$ is phonon density of states (DoS). To implement it, the phonon DoS is calculated from velocity auto-correlation of the MD trajectory, as detailed in the Supplemental Material and the works by Frenkel \cite{Frenkel2001} and Goddard \cite{Lin2003}.

The electronic entropy, the last term in Eqn. \ref{Equation:compoents}, is computed by averaging electronic entropy values of electronic structures obtained during the DFT-based MD trajectory. While our analysis presumes either diamagnetic or paramagnetic characteristics for the sake of simplicity, the underlying theoretical framework is sufficiently versatile to accommodate the examination of configurations influenced by magnetic spin orientations.

\begin{figure}[]
    \centering
    \includegraphics[width=0.49\textwidth]{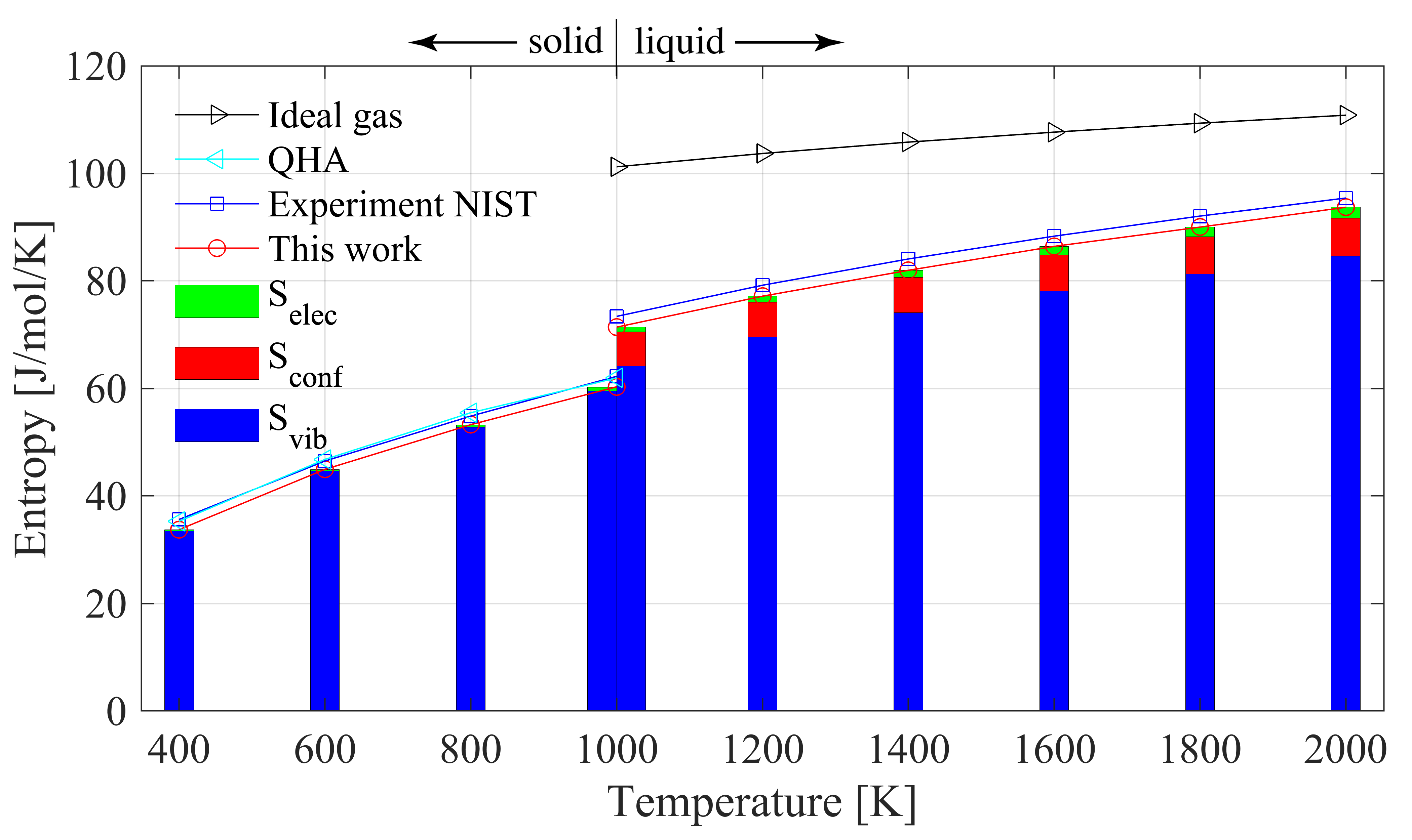} 
    \caption{Entropy and its configurational,  vibrational, and electronic components at various temperatures for solid (400-1000 K) and liquid (1000-2000 K) phases of aluminum.}
    \label{Figures:Al}
\end{figure}

The methodology presented herein exhibits four distinct advantages. First, it enables the direct computation of entropy from a singular MD trajectory, markedly simplifying the entropy calculation process. 
Second, it properly partitions coordinate space to configurational $S_{\text{conf}}$ and vibrational $S_{\text{vib}}$ components, followed by a summation of all states. This partitioning and summation are designed to (1) include all states significant to the system's entropy without missing any meaningful piece, and (2) preclude the possibility of state double-counting. These requirements had been the major challenge for direct entropy calculation, e.g., it is difficult to know the number of configurations in a liquid.
Third, the method ensures a highly accurate description of the vibrational entropy, often the predominant entropy term (see Fig. \ref{Figures:Al}). This accuracy is achieved through the evaluation of phonon DoS from velocity auto-correlation, an avenue that inherently accounts for quantum effects, thus offering an advantage over classical mechanics methods such as particle insertion or pairwise distribution analysis. While still under the harmonic approximation, the frequencies are derived from MD movements within the high-temperature region of the potential energy surface, thereby incorporating some degree of anharmonic effects. This approach contrasts with Hessian matrix calculation from atomic displacements, which is generally conducted at the bottom of the potential well. Finally, the proposed method demonstrates its versatility by generalizing across both solid and liquid states. This includes a wide spectrum of materials, ranging from defect-free ordered solids and partially disordered solids with defects to lattice-free liquids. As elucidated in the subsequent sections, this breadth underscores the method's applicability to a broad array of material phases and conditions.

\begin{table*}
\caption{\label{Table:results} Thermodynamic properties and transition dynamics of materials}
\centering
\begin{tabular}{l|l|r|rrrrr|r|rrr|rrr}
\hline
\multirow{2}{*}{Material} & \multirow{2}{*}{Phase} & \multicolumn{1}{|c|}{$T$ [K]} & \multicolumn{5}{c|}{$S$ [J/K/mol atom]} & \multicolumn{1}{c|}{$H$} & \multicolumn{1}{c}{$\Delta H$} & \multicolumn{1}{c}{$\Delta S$} & \multicolumn{1}{c|}{$T$} & $T_{\text{exp}}$ & $T_{\text{DFT}}$$^b$   \\
& & \hspace{.2cm}MD\hspace{.2cm} & \hspace{.1cm}Vib.\hspace{.1cm} & \hspace{.1cm}Conf.\hspace{.1cm} & \hspace{.1cm}Elec.\hspace{.1cm} & \hspace{.2cm}Total\hspace{.2cm} & \hspace{.2cm}Expt.$^a$\hspace{.2cm} &  \multicolumn{1}{c|}{[eV/atom]}& [J/mol at.] & [J/K/mol at.] & \multicolumn{1}{c|}{[K]} & [K] & [K]\\ \hline
Al& fcc & 1000 & 59.038 & 0.017 & 0.699 & 59.754 & 62.25$^c$ & -3.604 & \multicolumn{3}{l|}{fcc $\rightarrow$ liquid}\\
Al& liquid & 1000 & 63.807 & 6.339 & 0.870 & 71.016 & 73.40 & -3.502 & 9850 & 11.262 & 875 & 933 & 999  \\ \hline
Ti& bcc & 1800 & 79.451 & 2.369	& 7.189 & 89.009 & 87.47 & -7.435 & \multicolumn{3}{l|}{bcc $\rightarrow$ liquid}\\
Ti& liquid & 1800 & 81.461 & 6.424 & 7.050 & 94.935 & 94.42$^c$ & -7.336 & 10274 & 5.926 & 1734 & 1957 &1952\\ \hline
Si& diamond & 1400 & 57.604 & 0.000 & 0.178 & 57.781 & 56.46 & -5.218 & \multicolumn{3}{l|}{diamond $\rightarrow$ liquid}\\
Si& liquid & 1400 & 74.830 & 6.955 & 1.414 & 83.200 & 87.12$^c$ & -4.795 & 40767 & 25.418 & 1604 & 1687 & 1785\\ \hline
W& bcc & 3400 & 99.171 & 0.179 & 6.409 & 105.759 & 104.3 & -12.241 & \multicolumn{3}{l|}{bcc $\rightarrow$ liquid}\\
W& liquid & 3400 & 105.994 & 6.230 & 7.710 & 119.934 & 115.7$^c$ & -11.778 & 44745 & 14.175 & 3157 & 3685 & 3470 \\ \hline
Fe& bcc & 1500 & 66.912 & 3.274 & 6.600  & 76.786 & 84.52 & -7.790 & \multicolumn{3}{l|}{bcc $\rightarrow$ liquid}\\
Fe& liquid & 1500 & 70.428 & 6.167 & 7.167 & 83.762 & 92.9$^c$ & -7.657 & 12778 & 6.976 & 1832 & 1784 & 1828 \\ \hline
Fe& fcc & 1500 & 66.273 & 1.164 & 6.475 & 73.912 & 84.42 & -7.834 & \multicolumn{3}{l|}{fcc (metastable)$\rightarrow$liquid}\\
Fe& liquid & 1500 & 70.428 & 6.167 & 7.167 & 83.762 & 92.9$^c$ & -7.657 & 17094 & 9.850 & 1735 & - & - \\ \hline
Fe, 100GPa& hcp & 3500 & 77.627 & 0.947 & 11.158 & 89.732 & - & -1.729 & \multicolumn{3}{l|}{At 100GPa: hcp$\rightarrow$liquid}\\
Fe, 100GPa& liquid & 3500 & 81.571 & 6.039 & 12.003 & 99.613 & - & -1.369 & 34726 & 9.881 & 3514 & 3787 & 4019 \\ \hline
ZrO$_2$& fluorite & 2800 & 76.739 & 4.402 & 0.225 & 81.365 & 71.6 & -9.053 & \multicolumn{3}{l|}{fluorite $\rightarrow$ liquid} & \multicolumn{2}{l}{}\\
ZrO$_2$& liquid & 2800 & 77.739 & 8.118 & 0.285 & 86.142 & 81.3$^c$ & -8.907 & 14095 & 4.776 & 2951 & 2973 & - \\ \hline
HfO$_2$& fluorite & 3000 & 76.704 & 4.311 & 0.151 & 81.165 & - & -9.655 & \multicolumn{3}{l|}{fluorite $\rightarrow$ liquid} & \multicolumn{2}{l}{}\\
HfO$_2$& liquid & 3000 & 78.369 & 7.791 & 0.252 & 86.412 & - & -9.494 & 15577 & 5.246 & 2969 & 3053 & 3313 \\ \hline
Er$_2$O$_3$& H phase & 2900 & 82.413 & 4.455 & 0.209 & 87.077 & - & -7.841 & \multicolumn{3}{l|}{H phase $\rightarrow$ liquid} & \multicolumn{2}{l}{}\\
Er$_2$O$_3$& liquid & 2900 & 83.065 & 6.677 & 0.260 & 90.001 & - & -7.751 & 8786 & 2.924 & 3005 & 2693$^d$ & 2752 \\ \hline
HfC$_{0.5}$N$_{0.38}$& rocksalt & 4000 & 87.774 & 3.223 & 3.344 & 94.341 & - & -9.982 & \multicolumn{3}{l|}{rocksalt $\rightarrow$ liquid} & \multicolumn{2}{l}{}\\
HfC$_{0.5}$N$_{0.38}$& liquid & 4000 & 97.671 & 8.625 & 5.568 & 111.864 & - & -9.139 & 81333 & 17.523 & 4641 & - & 4141 \\ \hline
HfC$_{0.88}$ & rocksalt & 4000 & 87.764 & 2.890 & 3.595 & 94.249 & - & -9.726 & \multicolumn{3}{l|}{rocksalt $\rightarrow$ liquid} & \multicolumn{2}{l}{}\\
HfC$_{0.88}$ & liquid & 4000 & 97.791 & 8.717 & 5.818 & 112.327 & - & -8.904 & 79235 & 18.078 & 4383 & - & 3898 \\ \hline
HfC & rocksalt & 4000 & 86.173 & 2.219 & 3.046 & 91.438 & - & -9.719 & \multicolumn{3}{l|}{rocksalt $\rightarrow$ liquid} & \multicolumn{2}{l}{}\\
HfC & liquid & 4000 & 96.912 & 9.001 & 5.508 & 111.422 & - & -8.859 & 83013 & 19.984 & 4154 & 4223 & 3842 \\ \hline
\end{tabular}
\flushleft
$^a$ Experimental values from Ref. \onlinecite{nistchemistry}. $^b$ DFT values from SLUSCHI \cite{Hong2016}. $^c$ Metastable phase entropies extrapolated. $^d$ Ref. \cite{Ushakov2024Thermal}
\end{table*}



DFT calculations were performed by the Vienna Ab initio Simulation Package (VASP) \cite{VASP}, with the projector-augmented-wave (PAW) \cite{PAW} implementation and the generalized gradient approximation (GGA) for exchange correlation energy, in the form known as Perdew, Burke, and Ernzerhof (PBE) \cite{PBE}. 
The electronic temperature is accounted for by imposing a Fermi distribution of the electrons on the energy level density of states, so it is consistent with the ionic temperature. 
For all materials, the planewave energy cutoff is set to the default value (normal precision) of the pseudopotential during the MD simulations, and it is further increased to the high-precision value during the correction for the Pulay stress \cite{Francis1990}. 
DFT MD techniques are utilized to simulate atomic movements and trajectories. Specifically, MD simulations are carried out under constant number of atoms, pressure and temperature condition ($NPT$, isothermal-isobaric ensemble). Here the thermostat is conducted under the Nos\'{e}-Hoover chain formalism \cite{Nose1984,Nose1984-2,Hoover1985,Martyna1992}. The barostat is realized by adjusting volume every 80 steps according to average pressure, implemented in our \texttt{SLUSCHI} (Solid and Liquid in Ultra Small Coexistence with Hovering Interfaces) package \cite{Hong2016}. Although this does not formally generate an isobaric ensemble, this approach has been shown to provide an effective way to change volume smoothly and to avoid the unphysical large oscillation caused by commonly used barostats in small cells \cite{Hong2013}.

As our first example, we calculate the entropy of solid and liquid phase aluminum, comparing these findings with experimental data across a temperature range of 400 to 2000 K. Given aluminum's simple metallic nature and the high accuracy afforded by DFT, it exemplifies an ideal candidate for validating the efficacy of this approach. Such comparative analysis offers compelling validation for the precision of this new approach.

As shown in Fig. \ref{Figures:Al}, the new method exhibits remarkable accuracy in estimating the entropy of aluminum in both its solid and liquid states, with discrepancies limited to 3 J/mol/K. Further scrutiny into the entropy components reveals the method's significant advantages: the predominant contribution of vibrational entropy, compared with the considerably lesser role of configurational entropy, which is much smaller by an order of magnitude in liquids and is negligible in solids. This underscores the imperative of accurately quantifying vibrational entropy in thermodynamic assessments. By fundamentally prioritizing vibrational and phonon considerations, this new approach demonstrates a distinct superiority in capturing the intricate dynamics of entropy.

The success on both fcc Al, a defect-free solid with zero configurational entropy, and its liquid state, a lattice-free disordered phase with significantly higher configurational entropy, demonstrates the capability of this generalized approach in not only entropy calculations but also the determination of phase boundaries. Based on entropies calculated by this method and enthalpies from the same DFT MD trajectories, as shown in Table \ref{Table:results}, we are able to calculate the melting temperature of Al. The value of 875 K is in close agreement with the experimental melting temperature of 933 K.

The observed discrepancy, as illustrated in Figure \ref{Figures:Al}, can be attributed to the omission of long-wavelength phonons, a consequence of the finite dimensions of the simulation cell. This limitation results in an underrepresentation of these phonons, impacting the accuracy of our findings. To mitigate this source of error, two approaches are viable: expanding the size of the system under study, which would naturally incorporate a broader spectrum of phonon wavelengths, or applying a quadratic fit to the phonon density of states (DoS) curve in the low-frequency domain. The latter is predicated on the established principle that long-wavelength phonons exhibit behavior that can be closely approximated by quadratic fitting, offering a method to refine our model's precision in the future.


As our second example, we investigate zirconium oxide (ZrO$_2$), a material notable for its polymorphism and high-temperature stability \cite{Garvie1975,Fabrichnaya2007}. ZrO$_2$ undergoes a phase transformation to the fluorite structure at high temperatures before melting, which is characterized by a significant oxygen ion mobility \cite{kilo2003oxygen,wang2006zirconia}, due to oxygen sublattice melting. This property makes it an excellent candidate for applications as a fast ion conductor. However, the dynamic and partially disordered nature of oxygen diffusion at high temperatures presents a computational challenge for accurately determining the configurational entropy of the fluorite phase, as it assumes a non-zero value reflecting the multitude of attainable configurations. 

Our methodology addresses this complexity by enabling the calculation of both vibrational and configurational entropy components, thereby facilitating a comprehensive thermodynamic analysis of ZrO$_2$ in its high-temperature fluorite phase. As shown in Table \ref{Table:results}, the significant configurational entropy associated with oxygen (O: 6.574; Zr: 0.057 J/K/mol) underscores the fluidic nature of its sublattice, while zirconium atoms maintain their order. Combined with a calculation on liquid phase of ZrO$_2$, we determine the melting temperature at 2951 K, compared to the experimental value of 2973 K, and calculate the entropy of fusion to be 14.3 J/K/mol. This calculated fusion entropy, while seemingly lower than the NIST \cite{nistchemistry} reported value of 29 J/K/mol (which is derived from a linear extrapolation under a constant heat capacity assumption), is in fact more accurate. Contrary to the NIST extrapolation, our recent experimental and computational investigation of ZrO$_2$ yields an entropy of fusion of 17-18 J/K/mol \cite{Hong2018}, showcasing a reasonable agreement with the findings in the present work.


\begin{figure}[]
    \centering
    \includegraphics[width=0.49\textwidth]{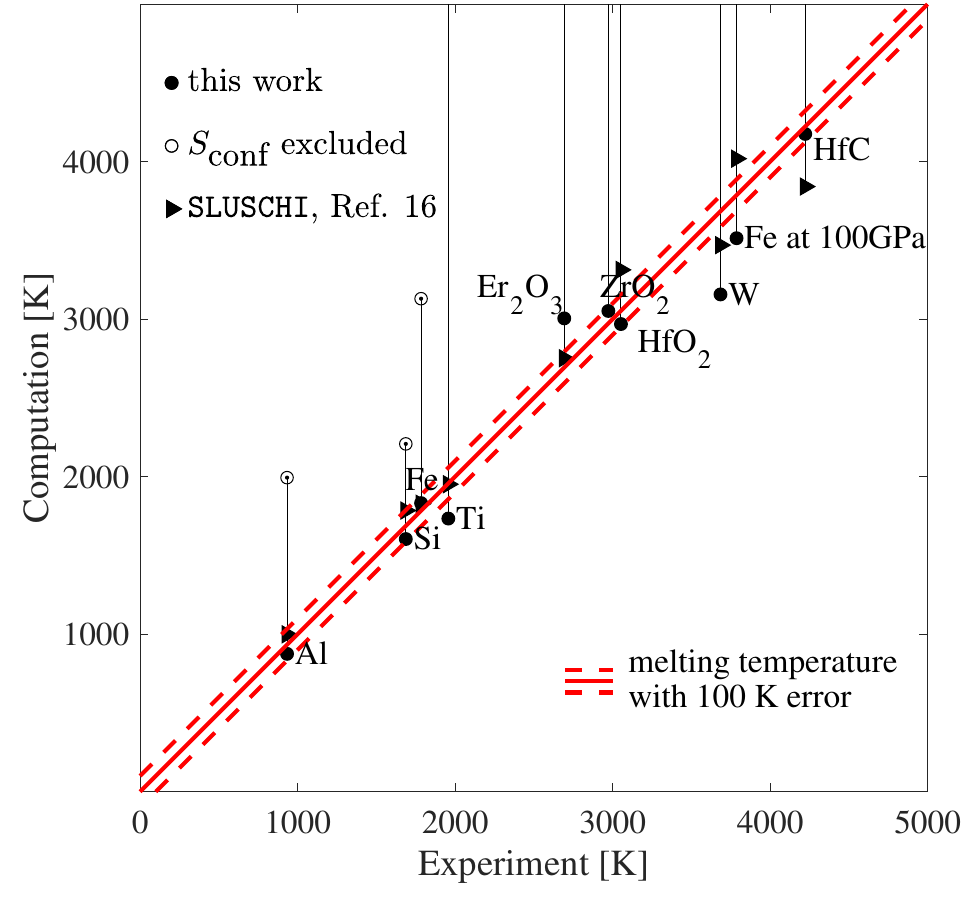} 
    \caption{Computational transition temperature based on this method versus experiments. Error bars of 100 K were plotted as red dash lines for eye guide. Plotted are melting temperature of materials at ambient pressure with the GGA-PBE functional employed unless elaborated in the label. Computational details can be found in the Supplemental Material. }
    \label{Figures:summary}
\end{figure} 

As our third example, we delve into hafnium carbonitride, the material we had previously identified \cite{HfCN2015} through DFT to surpass HfC and TaC and thereby set a new record for the highest melting temperature, prompting the synthesis and experimental investigation of the carbonitride materials systems \cite{Xie2023,Wyatt2023}. The computational prediction was later corroborated by experimental findings \cite{Buinevich2020}. Our new methodology now enables a comprehensive examination that extends to encompass melting temperature, fusion enthalpy, and critically, the entropy of this refractory material. Given the formula $T_m = \Delta H / \Delta S$, targeting materials with high fusion enthalpy or low fusion entropy emerges as a strategy for designing high melting temperatures. As revealed in Table \ref{Table:results}, the exceptionally large heat of fusion is a major contributing factor for the high melting temperature of the Hf-C and Hf-C-N systems, with 0.84 eV/atom for HfC$_{0.5}$N$_{0.38}$ compared to 0.82 for HfC$_{0.88}$ and 0.86 for HfC. However, an increase in fusion enthalpy alone does not justify the highest melting temperature of hafnium carbonitride: transitioning from HfC$_{0.88}$ to HfC$_{0.5}$N$_{0.38}$ yields only a 2.6\% increase in fusion enthalpy, whereas HfC exhibits the highest $\Delta H$ among them. Instead, our findings demonstrated that hafnium carbonitride derives thermodynamic stability from a significantly higher configurational entropy, primarily owing to vacancies and C/N disordering within its solid-state structure, with $\Delta S_{\textrm{conf}} = 5.40$ J/K/mol atom for HfC$_{0.5}$N$_{0.38}$ versus 5.83 for HfC$_{0.88}$ and 6.78 for HfC. Notably, these vacancies increase the configurational entropy of the solids, on the C/N sublattice, but not the liquids due to their absence of a lattice structure. The advancement in our methodology allows for the precise quantification of such effects, improving our understanding of the material's thermodynamic stability.

As our last example, we leverage the developed method to calculate melting temperatures, an important material property that necessitates extensive spatial sampling and thereby incurs significant computational costs \cite{Hongthesis}. Various materials are investigated, and the results are summarized in Fig. \ref{Figures:summary}.  This method demonstrates a high degree of accuracy, generally within a 100 K margin. This level of precision substantially affirms the reliability and effectiveness of our approach in computing entropy, applicable across various states of matter, including liquids, ordered solids, and solids exhibiting degrees of disorder.

The method presented exhibits certain limitations, particularly in its reliance on DFT MD simulations, which are constrained to exploring temporal spans of at most 100 picoseconds. This temporal limitation significantly restricts the diffusion of heavier atoms within ordered lattice structures. This limitation can be resolved in the future with atomic swapping under Metropolis Monte Carlo \cite{metropolis1953equation}. Additionally, the methodology's requirement for a clear segregation between configurational and vibrational entropy imposes constraints on the analysis. Specifically, it necessitates that vibrational states associated with adjacent configurations must be distinctly separate to preclude any potential for double counting. This aspect becomes particularly challenging in systems such as body-centered cubic titanium \cite{Petry1991,Kadkhodaei2017}, where the vibrational states of neighboring configurations inherently overlap, complicating the accurate partitioning of entropy contributions.

To summarize, we build a comprehensive methodology for the fast computation of entropy across both solid and liquid phases. The proposed method utilizes a single trajectory of MD to facilitate the calculation of entropy, including electronic, vibrational, and configurational entropies. We propose to calculate the configurational entropy, the most arduous component, as information entropy of nearest neighbors. We illustrate, through a variety of examples, that this method is both a versatile and valid technique for characterizing the thermodynamic states of both solids and liquids. Furthermore, this method is employed to expedite the calculation of melting temperatures, demonstrating its practical utility in computational thermodynamics. The method is implemented in our \texttt{SLUSCHI} package \cite{Hong2016} under the command \texttt{mds} (MD entropy).


\section*{Acknowledgements}
This research was supported by U.S. Department of Energy Office of Science under grant DE-SC0024724 and SC0023185, and by Arizona State University through the use of the facilities at its Research Computing.

\bibliography{citepapers}

\clearpage


\section{Supplemental Material}

\setcounter{equation}{0}
\setcounter{table}{0}
\setcounter{figure}{0}
\renewcommand{\theequation}{S\arabic{equation}}
\renewcommand{\thetable}{S\arabic{table}}
\renewcommand{\thefigure}{S\arabic{figure}}

\subsection{Boltzmann entropy vs. Gibbs entropy}

\subsubsection{Boltzmann entropy}
According to Boltzmann's statistical interpretation, the Helmholtz free energy is
\begin{equation}\label{Eqn:F}
F = -k_B T \ln{Z},
\end{equation}
where $Z$ is the partition function, a sum over all microstates $i$,
\begin{equation}
Z = \sum_i{\exp{\left(-\beta E_i\right)}},
\end{equation}
and 
\begin{equation}
\beta = \frac{1}{k_BT}.
\end{equation}

The differential form of Helmholtz free energy is 
\begin{equation}
dF = -pdV - SdT.
\end{equation}
Thus, 
\begin{equation}\label{Eqn:S}
S = - \left(\frac{\partial F}{\partial T}\right)_V
\end{equation}

Combine Eqns. \ref{Eqn:F} and \ref{Eqn:S},
\begin{eqnarray}\label{Eqn:Boltzmann}
S &=& k_B\ln{Z} + k_B T \cdot \frac{\sum_i{\left( \exp{\left(-\beta E_i\right)} \cdot E_i \cdot \frac{1}{k_B T^2} \right)}}{Z}\nonumber \\
&=& k_B\ln{Z} + \frac{\sum_i{p_i E_i}}{T} = \frac{- F + \langle E \rangle}{T},
\end{eqnarray}
where $p_i$ is the probability of microstate $i$,
\begin{equation}
p_i = \frac{\exp{\left(-\beta E_i\right)}}{Z},
\end{equation}
and the probabilities $p_i$ is normalized
\begin{equation}
\sum_{i}{ p_i }=1.
\end{equation}

\subsubsection{Gibbs entropy}
Gibbs generalized the interpretation of entropy as 
\begin{equation}
S = -k_B \sum_i{ \left( p_i \cdot \ln{p_i} \right) }.
\end{equation}

It is straightforward to prove that this entropy $S$ is equivalent to the Boltzmann entropy.
\begin{eqnarray}\label{Eqn:Gibbs}
S &=& -k_B \sum_i{ \left( p_i \cdot \ln{p_i} \right) } \nonumber \\
&=& -k_B \sum_i{ \left( p_i \cdot \ln{ \left(\frac{\exp{\left(-\beta E_i\right)}}{Z}\right)} \right) } \nonumber \\
&=& -k_B \sum_i{ \left( p_i \cdot \left( -\beta E_i - \ln{Z} \right) \right)}\nonumber \\
&=& \frac{\sum_i{p_i E_i}}{T} + k_B\ln{Z}.
\end{eqnarray}
Now it is clear that Eqns \ref{Eqn:Boltzmann} (Boltzmann) and \ref{Eqn:Gibbs} (Gibbs) are the same and the Boltzmann and Gibbs equations can be used interchangeably.

\subsection{Segmentation of $S_{\text{conf}}$ and $S_{\text{vib}}$}
We then perform the segmentaion of $S$ into $S_{\text{conf}}$ and $S_{\text{vib}}$, by grouping a microstate $i$ into the configuration $\sigma_j$ it belongs to, as shown in Fig. \ref{Fig:diagram}. According to Gibbs entropy,
\begin{eqnarray}\label{Eqn:segment}
S &=& -k_B \sum_i{ \left( p_i \cdot \ln{p_i} \right) } \nonumber \\
&=& -k_B \sum_j{ \sum_{i \textrm{ in } \sigma_j}\left( p_i \cdot \ln{p_i} \right) } \nonumber \\
&=& -k_B \sum_j{ \sum_{i \textrm{ in } \sigma_j}\left( p_i \cdot \ln{\left(\tilde{p}_{\sigma_j} \cdot \frac{p_i}{\tilde{p}_{\sigma_j}}\right)} \right) } \nonumber \\
&=& -k_B \sum_j{ \left( \tilde{p}_{\sigma_j} \cdot \ln{\tilde{p}_{\sigma_j}} \right) } \nonumber \\
&& -k_B \sum_j{ \sum_{i \textrm{ in } \sigma_j} \left( p_i \cdot \ln{\frac{p_i}{\tilde{p}_{\sigma_j}}} \right) },
\end{eqnarray}
where $\tilde{p}_{\sigma_j}$ is the sum of $p_i$ in configuration $\sigma_j$ and thus the probability of the configuration,
\begin{equation}
\tilde{p}_{\sigma_j} = \sum_{i \textrm{ in } \sigma_j}{ p_i },
\end{equation}
and the probabilities $\tilde{p}_{\sigma_j}$ are normalized due to $p_i$ being normalized,
\begin{equation}
\sum_{j}{ \tilde{p}_{\sigma_j} }=1.
\end{equation}

We interpret the first term in Eqn. \ref{Eqn:segment} as the configurational entropy $S_{\text{conf}}$, and we employ the formula to compute its value.
\begin{equation}\label{Eqn:Sconf}
S_{\text{conf}} = -k_B \sum_j{ \left( \tilde{p}_{\sigma_j} \cdot \ln{\tilde{p}_{\sigma_j}} \right) }
\end{equation}
For the second term in Eqn. \ref{Eqn:segment}, it is clear that it is the vibrational entropy $S_{\text{vib}}$.
\begin{eqnarray}\label{Eqn:Svib}
&& -k_B \sum_j{ \sum_{i \textrm{ in } \sigma_j} \left( p_i \cdot \ln{\frac{p_i}{\tilde{p}_{\sigma_j}}} \right) } \nonumber\\
&=& -k_B \sum_j{ \sum_{i \textrm{ in } \sigma_j} \left( \tilde{p}_{\sigma_j} \cdot \frac{p_i}{\tilde{p}_{\sigma_j}} \cdot \ln{\frac{p_i}{\tilde{p}_{\sigma_j}}} \right) } \nonumber\\
&=& \sum_j{ \tilde{p}_{\sigma_j} \cdot \left( -k_B  \sum_{i \textrm{ in } \sigma_j} \left( \frac{p_i}{\tilde{p}_{\sigma_j}} \cdot \ln{\frac{p_i}{\tilde{p}_{\sigma_j}}} \right) \right) } \nonumber \\
&=& \sum_j{ \tilde{p}_{\sigma_j} \cdot S_{\text{vib},\sigma_j} } = \langle S_{\text{vib}} \rangle_{\sigma} = S_{\text{vib}}.
\end{eqnarray}
For the last two lines, when we focus on configuration $\sigma_j$,
\begin{equation}
S_{\text{vib},\sigma_j} = S_{\sigma_j} =  -k_B  \sum_{i \textrm{ in } \sigma_j} \left( \frac{p_i}{\tilde{p}_{\sigma_j}} \cdot \ln{\frac{p_i}{\tilde{p}_{\sigma_j}}} \right) ,
\end{equation}
where $p_i/\tilde{p}_{\sigma_j}$ is the conditional probability $p(i | \sigma_j)$ and it is normalized. We calculate the ensemble average of the vibrational entropy Eqn. \ref{Eqn:Svib} from velocity auto-correlation in molecular dynamics trajectories, as detailed in the works by Frenkel \cite{Frenkel2001} and Goddard \cite{Lin2003}.


\subsection{Configurational entropy $S_{\text{conf}}$}
Eqn. \ref{Eqn:Sconf} provides us an approach to evaluate the configurational entropy. To achieve this, we need to compute the probability $\tilde{p}_{\sigma}$ of each configuration $\sigma$. Here we describe a configuration by the number of nearest neighbors. Denote $N$ as the number of atoms and $\sigma(n_1,n_2,...,n_N)$ as the configuration, where $n_k$ is the number of nearest neighbors of the $k$th atom. The probability is 
\begin{eqnarray}\label{Eqn:independent_assumption}
\tilde{p}_{\sigma(n_1,n_2,...,n_N)} = \Pi_{k=1}^N P_{n_k}\cdot \frac{\tilde{p}_{\sigma(n_1,n_2,...,n_N)}}{\Pi_{k=1}^NP_{n_k}},
\end{eqnarray}
where $P_{n}$ stands for the overall probability of a single atom with $n$ nearest neighbors.
\begin{equation}
\sum_{n}{ P_{n} }=1.
\end{equation}
The ratio $\tilde{p}_{\sigma(n_1,n_2,...,n_N)}/\Pi_{k=1}^NP_{n_k}$ is the correlation between the probability of an $N$-atom configuration and the product of probabilities of all its individual atoms. Now let us assume the ratio is 1, i.e., the atoms are independent. Thus, Eqn. \ref{Eqn:independent_assumption} becomes 
\begin{eqnarray}\label{Eqn:product}
\tilde{p}_{\sigma(n_1,n_2,...,n_N)} = \Pi_{k=1}^N P_{n_k}.
\end{eqnarray}
Plug it into Eqn. \ref{Eqn:Sconf},
\begin{eqnarray}\label{Eqn:derive}
S_{\text{conf}} &=& -k_B \sum_{n_1,n_2,...,n_N}{ \left( \Pi_{k=1}^N P_{n_k} \cdot \ln{\left( \Pi_{k=1}^N P_{n_k} \right)} \right) } \nonumber \\
&=& -k_B \sum_{n_1,n_2,...,n_N}{ \left( \Pi_{k'=1}^N P_{n_{k'}} \cdot \sum_{k=1}^N \ln{P_{n_k}} \right) } \nonumber \\
&=& -k_B \sum_{n_1,n_2,...,n_N}{ \left( \sum_{k=1}^N \left( \Pi_{k'=1}^N P_{n_{k'}} \cdot \ln{P_{n_k}} \right) \right)} \nonumber \\
&=& -k_B N \sum_{n_1,n_2,...,n_N}{ \left( \Pi_{k'=1}^N P_{n_{k'}} \cdot \ln{P_{n_k}} \right) } \nonumber \\
&=& -k_B N \sum_{n_k}{ \left( P_{n_{k}} \cdot \ln{P_{n_k}} \right) } \nonumber \\
&=& -k_B N \sum_{n}{ \left( P_{n} \cdot \ln{P_{n}} \right) } .
\end{eqnarray}
The configurational entropy per atom is 
\begin{equation}
S_{\text{conf}} = -k_B \sum_{n}{ \left( P_{n} \cdot \ln{P_{n}} \right) }.
\end{equation}
Because nearest neighbors are pairwise, the sum of $n_k$ must be an even number, when the pair involves the same element.
\begin{equation}
\frac{\tilde{p}_{\sigma(n_1,n_2,...,n_N)}}{\Pi_{k=1}^NP_{n_k}} = \begin{cases} 
  1 & \text{if $\sum_{k=1}^N n_k$ is even}, \\
  0 & \text{if $\sum_{k=1}^N n_k$ is odd}.
\end{cases}
\end{equation}
This effectively reduces the entropy by half. Therefore, 
\begin{equation}\label{Eqn:final_Svib}
S_{\text{conf}} = -\left(1-\frac{\delta}{2}\right)k_B \sum_{n}{ \left( P_{n} \cdot \ln{P_{n}} \right) },
\end{equation}
where $\delta$ assigned a value of 1 if the the pair consists of the same element, and 0 otherwise.

In Eqn. \ref{Eqn:final_Svib} we have simplified the configurational entropy to a formula we can readily compute. $P_{n}$ is the probability of a local atom having $n$ nearest neighbors, which we can evaluate from an MD trajectory.

\subsection{The density of state function}

Phonons represent the quantized collective oscillations of atoms within a solid, and their density of states (DoS) can be modeled as the distribution of vibrational normal modes within a sufficiently large supercell. The density of these modes, i.e., effective vibration intensity, is derived from aggregating the contributions across all atoms contained within the supercell,
\begin{equation}\label{Eqn:Svib}
g(\nu) = \frac{2}{k_BT} \sum_{j=1}^{N} \sum_{k=1}^{3} m_j s^j_k(\nu),
\end{equation}
where \( m_j \) is the mass of atom \( j \). The spectral density \( s^j_k(\nu) \) for atom \( j \) along the \( k \)th coordinate (\( k=x, y, \) and \( z \) in Cartesian coordinates) is computed through the square of the Fourier transform of its velocity components,
\begin{equation}
s^j_k(\nu) = \lim_{\tau \to \infty} \frac{\left| \int_{-\tau}^{\tau} v^j_k(t) e^{-2\pi i \nu t} dt \right|^2}{2\tau} = \lim_{\tau \to \infty} \frac{1}{2\tau} \left| A^j_k(\nu) \right|^2,
\end{equation}
where \( v^j_k(t) \) is the \( k \)th velocity component of atom \( j \) at time \( t \), and
\begin{equation}
A^j_k(\nu) = \lim_{\tau \to \infty} \int_{-\tau}^{\tau} v^j_k(t) e^{-2\pi i \nu t} dt.
\end{equation}

The density of states function can also be derived using the Fourier transform of the velocity autocorrelation function. The total velocity autocorrelation function, denoted as \( C(t) \), is defined as the mass-weighted aggregate of the velocity autocorrelation functions of individual atoms,
\begin{equation}
C(t) = \sum_{j=1}^{N} \sum_{k=1}^{3} m_j c^j_k(t),
\end{equation}
where \( c^j_k(t) \) is the velocity autocorrelation for atom \( j \) along the \( k \) direction
\begin{eqnarray}
c^j_k(t) &=& \lim_{\tau \to \infty} \frac{\int_{-\tau}^{\tau} v^j_k(t') v^j_k(t'+t) dt'}{\int_{-\tau}^{\tau} dt'} \nonumber \\
&=& \lim_{\tau \to \infty} \frac{1}{2\tau} \int_{-\tau}^{\tau} v^j_k(t') v^j_k(t'+t) dt'.
\end{eqnarray}
In accordance with the Wiener–Khinchin theorem, the atomic spectral density \( s^j_k(\nu) \) is directly determined by performing the Fourier transform on \( c^j_k(t) \),
\begin{equation}
s^j_k(\nu) = \lim_{\tau \to \infty} \frac{\left| \int_{-\tau}^{\tau} c^j_k(t) e^{-2\pi i \nu t} dt \right|^2}{2\pi}.
\end{equation}

Therefore, \( S(\nu) \) as outlined in Eqn. \ref{Eqn:Svib} can be derived through the Fourier transform of \( C(t) \),
\begin{eqnarray}
g(\nu) &=& \frac{2}{k_BT} \sum_{j=1}^{N} \sum_{k=1}^{3} m_j s^j_k(\nu) \nonumber \\
&=& \frac{2}{k_BT} \lim_{\tau \to \infty} \int_{-\tau}^{\tau} \left( \sum_{j=1}^{N} \sum_{k=1}^{3} m_j c^j_k(t) \right) e^{-2\pi i \nu t} dt \nonumber \\
&=& \frac{2}{k_BT} \lim_{\tau \to \infty} \int_{-\tau}^{\tau} C(t) e^{-2\pi i \nu t} dt.
\end{eqnarray}

\subsection{Thermodynamic properties\\ from phonon density of states}

The energy for a quantum harmonic oscillator is expressed as
\begin{equation}
\epsilon = (n+\frac{1}{2})h \nu, \nonumber \\
\end{equation}
where $n$ denotes the quantum number, and $\nu$ the vibrational frequency.

The partition function is then given by
\begin{eqnarray}
Z &=& \sum_{n=0}^\infty{\exp(-\frac{\epsilon}{k_BT})} \nonumber  \\
&=& \sum_{n=0}^\infty{\exp(-\frac{(n+\frac{1}{2})h \nu}{k_BT})} \nonumber \\
&=& \exp(-\frac{h \nu}{2k_BT})\sum_{n=0}^\infty{\exp(-\frac{nh \nu}{k_BT})} \nonumber \\
&=& \exp(-\frac{h \nu}{2k_BT})\sum_{n=0}^\infty{\left[\exp(-\frac{h \nu}{k_BT})\right]^n} \nonumber \\
&=& \exp(-\frac{h \nu}{2k_BT}) \frac{1}{1-\exp(-\frac{h \nu}{k_BT})} \nonumber \\
&=& \frac{1}{\exp(\frac{h \nu}{2k_BT})-\exp(-\frac{h \nu}{2k_BT})} \nonumber \\
&=& \frac{1}{2\sinh(\frac{h \nu}{2k_BT})}
\end{eqnarray}

Subsequently, the Helmholtz free energy is derived as
\begin{eqnarray}
F &=& -k_BT\ln{Z} \nonumber \\
&=& k_BT \ln{\left[2\sinh(\frac{h \nu}{2k_BT})\right]}.
\end{eqnarray}

For a solid characterized by a phonon density of states $g(\nu)$, the Helmholtz free energy is extended to
\begin{equation}
F = k_BT\int{ \ln{\left[2\sinh(\frac{h \nu}{2k_BT})\right]}g(\nu)d\nu},
\end{equation}
as in Eqn. \ref{Eqn:Svib_solid}. This integral form accounts for the contributions of all phonon modes to the solid's thermodynamic properties.


\end{document}